\documentclass[aps, prb, twocolumn, superscriptaddress, showpacs]{revtex4-2}
\usepackage{color, graphicx}
\usepackage{amssymb,amsmath, amsfonts, bm}
\usepackage[colorlinks,citecolor=blue, urlcolor=blue, linkcolor=blue]{hyperref}
\usepackage{bbold}

\usepackage{newtxtext,newtxmath}
\usepackage{mathtools}
\usepackage{dcolumn}
\usepackage[none]{hyphenat} 
\usepackage{tabularx}

\begin{document}
\title{Reentrance of metal-insulator transition and magnetic competitions on a triangular lattice with second nearest-neighbor hopping}  

\author{Xin Gao}
\author{Cong Hu}
\affiliation{School of Physical Science and Technology, ShanghaiTech University, Shanghai 201210, China}
\affiliation{Shanghai Institute of Optics and Fine Mechanics, Chinese Academy of Sciences, Shanghai 201800, China}
\affiliation{University of Chinese Academy of Sciences, Beijing 100049, China}
\author{Jian Sun}
\affiliation{School of Physical Science and Technology, ShanghaiTech University, Shanghai 201210, China}
\author{Xiao-Qun Wang}
\affiliation{Department of Physics and Astronomy, Shanghai Jiao Tong University, 800 Dongchuan Road, Shanghai 200240, China}
\affiliation{Key Laboratory of Artificial Structures and Quantum Control of MOE, Shenyang National Laboratory for Materials Science, Shenyang 110016, China}
\affiliation{Beijing Computational Science Research Center, Beijing 100193, China}
\author{Hai-Qing Lin}
\affiliation{Beijing Computational Science Research Center, Beijing 100193, China}
\author{Gang Li}
\email{ligang@shanghaitech.edu.cn}
\affiliation{School of Physical Science and Technology, ShanghaiTech University, Shanghai 201210, China}

\pacs{71.10.Fd, 71.27.+a, 71.30.+h}

\begin{abstract}
The $120^{\circ}$ antiferromagnetism (AFM) is widely believed as the magnetic ground state of the triangular systems because of the geometrical frustration. 
The emergence of novel magnetism, such as the row-wise AFM in Mn/Cu(111) and Sn/Si(111), reveals the importance of the longer-range hopping on magnetic competitions in realistic material systems. 
By utilizing advanced many-body techniques, we systematically studied the isotropic triangular Hubbard model with second nearest-neighbor hopping $t^{\prime}$, including both the single- and the two-particle responses. 
We found that both electronic and magnetic phase transitions show a clear dependence on $t^{\prime}/t$. 
Consequently, we observed a remarkable reentrance of the metal-insulator transition and a crossover between the $120^{\circ}$- and the row-wise AFM.
The Fermi surface (FS) shows two distinct structures with the nesting vectors consistent with the magnetic correlations. 
When $t^{\prime}$ evolves from 0 to 1, the correlated Fermi surface demonstrates a Lifschitz transition between the two nesting structures, and exotic phases like the featureless insulating state can be realized. 
Our work sheds light on the engineering of electronic and magnetic correlations of correlated triangular surfaces via longer-range hopping.
The rich phase diagram and the high degree of tunability make the triangular lattice with longer-range hopping a more realistic platform to study the emergent magnetic competitions. 
\end{abstract}
\maketitle

\section{introduction}
Strongly correlated electron systems are in heart of contemporary research of condensed matter physics. 
Among them, the two dimensional (2D) electron systems, in which the quantum wave function of electrons is confined in 2D layer, exhibit distinct and diverse phenomena continuously challenging our understanding of quantum physics. 
Examples include the quantum Hall systems~\cite{PhysRevLett.45.494, doi:10.1143/JPSJ.39.279}, quantum wells~\cite{Koenig02112007, PhysRevLett.106.126803}, orbital-selective Mott insulators~\cite{PhysRevLett.92.216402, PhysRevB.72.205124, PhysRevLett.102.126401}, spin liquid materials~\cite{ANDERSON1973153, ANDERSON1196}, Hund's coupled iron-based superconductors~\cite{Kamihara:2006aa, Kamihara:2008aa, Chen:2008aa, B813153H, PhysRevLett.101.107006}, transition-metal oxide heterostructures~\cite{Ohtomo:2004aa, Miao:2020aa}, and the recently discovered infinite-layer nickel oxides~\cite{Li:2019aa} and twisted bilayer graphene~\cite{Cao:2018aa, Stepanov:2020aa, Jiang:2019aa, Lu:2019aa, Sharpe605, Xie:2019aa, Yankowitz1059} etc.   
Among the 2D systems, the triangular system is unique. In addition to the spatial confinement, the geometric frustration makes the spin arrangement a nontrivial problem. 
Electron spins on a triangular lattice can be neither parallel nor antiparallel simultaneously with respect to all other neighboring spins. 
A compromise has to be made, which leads to an arrangement of spins with 120$^{\circ}$ angle pointing from one to another in the classical ground state.
The noncollinear antiferromagnetism (AFM) in triangular lattice is a natural consequence of the geometric frustration and has been found in many systems, such as Cr/Fe(111)~\cite{0953-8984-11-48-302},  Mn/Ag(111)~\cite{PhysRevLett.101.267205}, Cr/Cu(111)~\cite{0953-8984-11-48-302, PhysRevLett.86.1106}.

The spin arrangements in triangular materials can be much more prosperous.  
The spin frustration and, in particular, its competition with strong electronic correlations triggers the emergence of various unconventional phenomena in triangular systems, including the possibility of the disordered ground state to occur in $S=\frac{1}{2}$ 2D AFM~\cite{0034-4885-74-5-056501, nature_balents}. 
A typical example is the triangular-layered organic $\kappa\mbox{-(BEDT-TTF)}_{2}\mbox{Cu}_{2}\mbox{(CN)}_{3}$, whose bulk spin susceptibility~\cite{PhysRevLett.91.107001} shows no
indication of long-range AFM order at significantly
lower temperature as compared to the Heisenberg exchange theoretically estimated from the high-temperature series
expansion~\cite{PhysRevB.71.134422}.  
In sharp contrast to the 120$^{\circ}$ noncollinear AFM, the appearance of nonmagnetic quantum spin-liquid phase in $\kappa\mbox{-(BEDT-TTF)}_{2}\mbox{Cu}_{2}\mbox{(CN)}_{3}$ is  astonishing.   
Many theoretical works have been devoted to this challenging problem~\cite{PhysRevB.72.045105, PhysRevB.60.1064, doi:10.1143/JPSJ.71.2109, PhysRevLett.95.036403, PhysRevLett.97.046402, PhysRevLett.98.067006, PhysRevLett.115.167203, PhysRevB.95.165110, luo2019gapless, Kira2020}, with the conclusions converging to the competition of geometrical frustration and electronic correlations. 

Besides the spin-liquid state, the stabilization of a collinear AFM in correlated triangular lattice is another surprise, which has been confirmed theoretically in Mn/Cu(111)~\cite{PhysRevLett.86.1106} and experimentally in Sn/Si(111)~\cite{nat.comm_2013,PhysRevB.90.125439} surfaces.
Different from the normal triangular lattice, the spin susceptibility of these systems peaks at $\mathbf{M}$-point instead of at $\mathbf{K}$ indicating a row-wise-type collinear AFM. 
The clear contradiction to the classical spin arrangement calls for new understandings of correlated magnetism in quantum triangular systems. 
In Sn/Si(111), in addition to the nearest-neighbor (N.N.) hopping generally considered in various quantum many-body model studies, there exists considerably large second N.N. hopping, providing a new competing energy scale to the geometrical frustration. 
The relationship of the row-wise collinear AFM with the second N.N. hopping, especially under strong electronic correlation, has not been fully explored. 
By using the dual-fermion (DF) approach~\cite{PhysRevB.77.033101, PhysRevB.79.045133}, we  have studied the magnetic correlations in an effective model for Sn/Si(111) and explained the collinear AFM observed experimentally~\cite{nat.comm_2013,PhysRevB.90.125439}.
Recently, K. Misumi et al.~\cite{PhysRevB.95.075124} also studied this problem by using zero-temperature variational cluster approximation~\cite{refId0, PhysRevLett.91.206402}, which reaches a consistent conclusion. 

In this work, we want to present a systematic study of the isotropic triangular lattice as a function of the second N.N. hopping $t^{\prime}/t$. 
We find a strong competition of $120^{\circ}$- and row-wise collinear AFM, and a reentrance of metal-insulator transition (MIT), which significantly enrich our understanding of the correlated triangular materials.  

\section{model and method}
\label{Sec:model_method}
We study the isotropic triangular lattice by considering the following Hubbard model at half-filling,
\begin{equation}\label{Ham}
H = -t\sum_{\langle i,j\rangle}(c_{i\sigma}^{\dagger}c_{j\sigma}+h.c.)-t^{\prime}\sum_{[i,j]}(c_{i\sigma}^{\dagger}c_{j\sigma}+h.c.) + U\sum_{i}n_{i\uparrow}n_{i\downarrow}\:.
\end{equation}
To resemble the realistic triangular material systems, in addition to the N.N. hopping $t$ between $\langle i, j\rangle$, we further include the $t^{\prime}$ term between the pair of the second N.N. sites $[i, j]$.   
Compared to the ideal triangular Hubbard model with only N.N. hopping, the presence of this longer-range hopping term delocalizes the electrons and further competes with the geometrical frustration. 
The local Coulomb interaction between two electrons with opposite spins from the same site is $U$. 
Throughout the paper, we take the energy unit $t$ to be 1. Whenever $t^{\prime}$ and $U$ are referred to, they shall be understood as $t^{\prime}/t$ and $U/t$.

\begin{figure}[htpb]
\centering
\includegraphics[width=\linewidth]{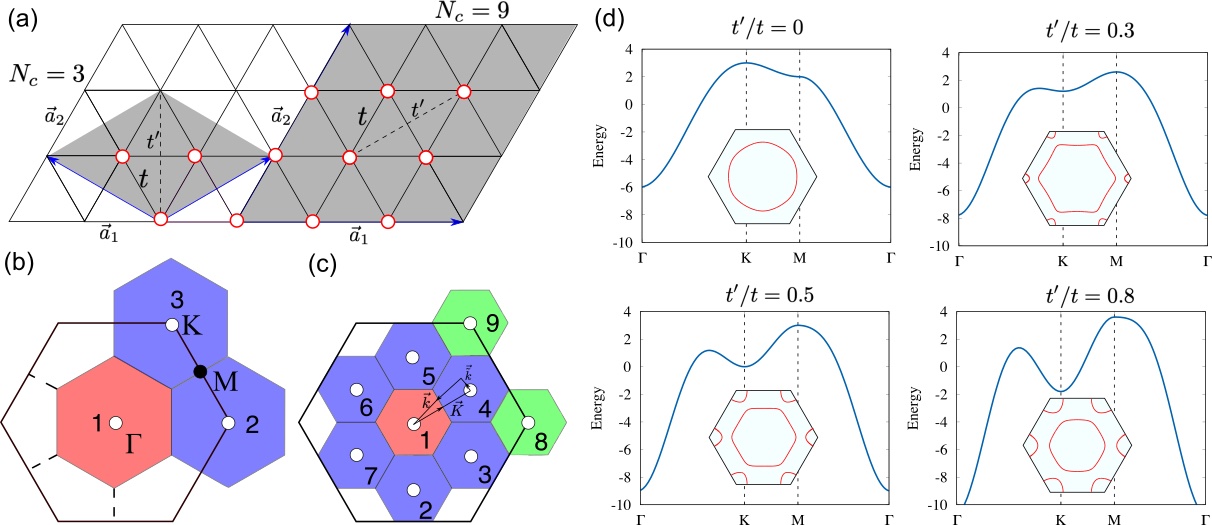}
\caption{(a) Two DCA clusters with $N_{c}=3$ and $N_{c}=9$ sites and (b), (c) with their corresponding BZ. The color shaded areas in (a) display the primitive cell of the two clusters with the 3 and 9 sites denoted by the red circles. In (b) and (c), we show the momentum patches in different colors. All patches with the same color belong to the same symmetry group. The white circle inside each momentum patch corresponds to the center momentum of this patch.
(d) The band dispersion and Fermi surface topology of four representative $t^{\prime} $.}
\label{Fig:cluster_BZ}
\end{figure}

To better account for the competition and the interplay between $t$ and $t^{\prime}$, we employ a self-consistent method which essentially works at thermodynamic limit and respects the periodicity of the Brillouin zone (BZ), i.e., the dynamical cluster approach (DCA)~\cite{RevModPhys.77.1027}. 
In this work, we consider a $N_{c}=9$ site cluster whose BZ is patched into nine sections as shown in Fig.~\ref{Fig:cluster_BZ}.
The specific shape of a finite-size cluster can play a crucial role in the calculations, which may either break or additionally impose symmetries into the calculations. DCA employs the periodic boundary condition, which effectively restores the translational symmetry. The BZ of the Nc = 9 site cluster chosen in our calculations respects all the lattice symmetries including the six-fold rotational and mirror symmetries, leading to the titling of momentum patches in the BZ shown in Fig. 1(c). We have carefully verified and further ensured these symmetries to be satisfied in every DCA iteration.
Not all nine patches give independent self-energy functions. 
Under six-fold rotation  and mirror symmetries, some momentum patches become equivalent. In Fig.~\ref{Fig:cluster_BZ}(c) we show all the equivalent momentum patches with the same color. 
In a $N_{c}=9$ cluster, there are only three inequivalent momentum patches, providing three independent self-energy functions. 
As a comparison, we also showed a $N_{c}=3$ site cluster in Fig.~\ref{Fig:cluster_BZ}(a), whose BZ is divided into two inequivalent patches under the DCA construction. 
We note that the study of the two different magnetic correlations requires a resolution of the self energy at two inequivalent momentum points $\mathbf{K}$ and $\mathbf{M}$. 
As clearly seen in Fig.~\ref{Fig:cluster_BZ}(b), $\mathbf{K}$ and $\mathbf{M}$ reside in the same momentum patch. 
Consequently, the self-energy at these two points will be exactly same in $N_{c}=3$ DCA calculations. 
For this reason and also for better accounting for the nonlocal effect, we adopt the $N_{c}=9$ site cluster. 

To solve the DCA self-consistent equation, we adopt the interaction-expansion CT-QMC method~\cite{PhysRevB.72.035122, RevModPhys.83.349}, 
and measure the single particle Green's function directly in the Matsubara frequency space. 
To monitor the instability of the magnetic channel, we also measure the particle-hole vertex function in the last iteration after the DMFT self-consistency is achieved. 

\section{Metal-Insulator Transition}
\label{Sec:MIT}
The MIT in isotropic triangular lattice has been widely studied by various theoretical methods~\cite{doi:10.1143/JPSJ.71.2109, PhysRevLett.97.046402, doi:10.1143/JPSJ.75.074707, doi:10.1143/JPSJ.76.074719, PhysRevB.78.205117,  PhysRevB.77.214505, PhysRevLett.100.136402,  PhysRevLett.100.076402, PhysRevLett.101.166403, PhysRevB.79.115116, PhysRevB.79.195108, PhysRevLett.105.267204, PhysRevB.86.235137, PhysRevLett.110.206402,PhysRevB.87.035143, PhysRevB.89.195108,  PhysRevB.89.235107,PhysRevB.89.161118, PhysRevB.91.245125, PhysRevResearch.2.013295}. 
Single-site DMFT correctly captures the essence of paramagnetic MIT in low-dimensional systems, but with the incorrect estimation of the transition boundary.
In particular, it predicts an increasing critical $U_{c}$ for MIT at lower temperature on 2D square lattice.
The cellular DMFT~\cite{PhysRevLett.87.186401, RevModPhys.78.865}, on the other hand, revealed a decreasing $U_{c}$ with the decrease of temperature~\cite{PhysRevLett.101.186403}. 
The difference shows the significant role played by the nonlocal charge fluctuations. 
Compared to the square lattice, a local approximation like DMFT is better justified in triangular lattice.  
Although the triangular lattice is often taken as a prototype of frustrated systems where the local fluctuations dominate, the nonlocal correlation effect is not negligible.
Thus, it is not yet fully settled, in isotropic triangular lattice with only N.N. hopping $t$, how the MIT behaves as a function of temperature under the nonlocal charge fluctuations included in the cluster type DMFT calculations. 
\begin{figure}[htbp]
\centering
\includegraphics[width=0.9\linewidth]{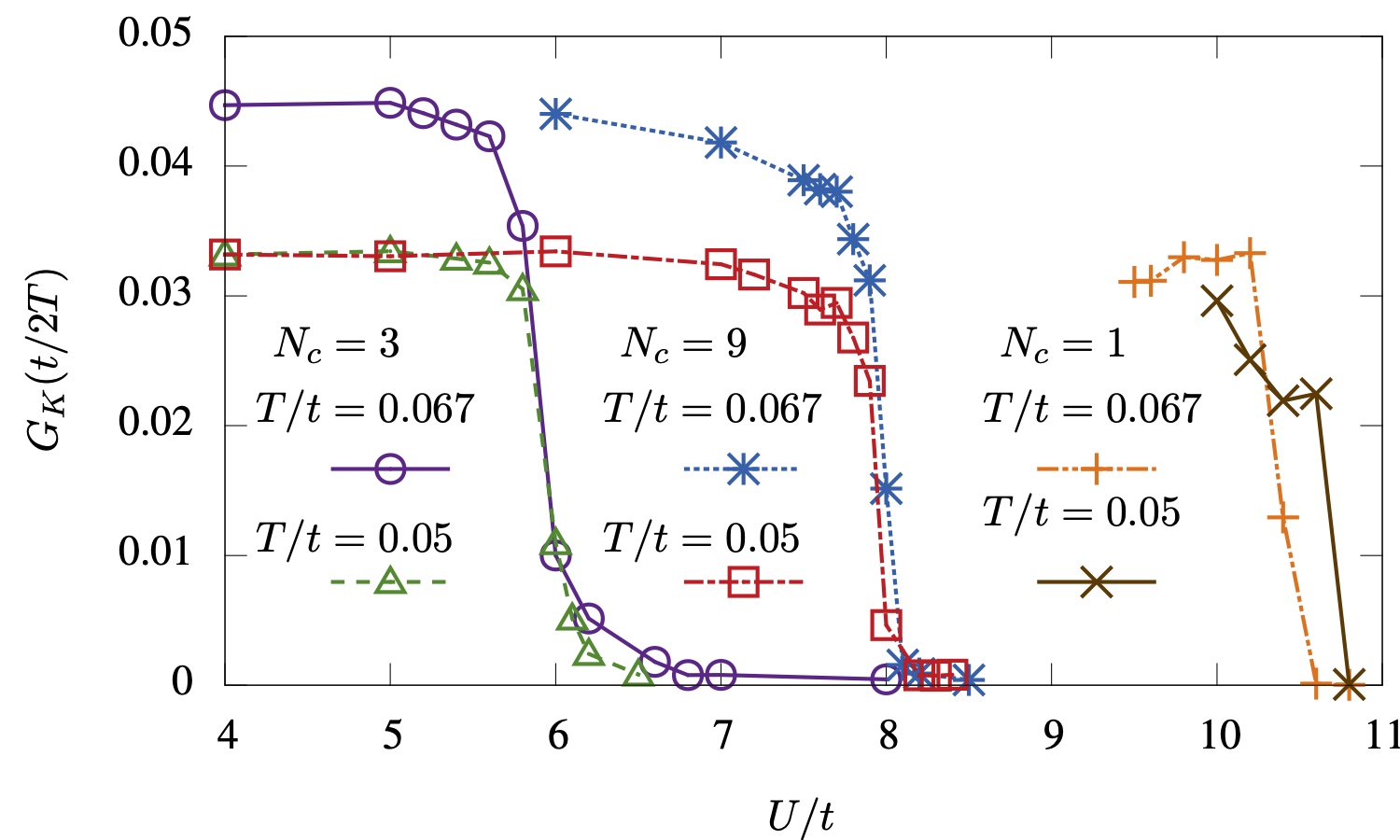}
\includegraphics[width=\linewidth]{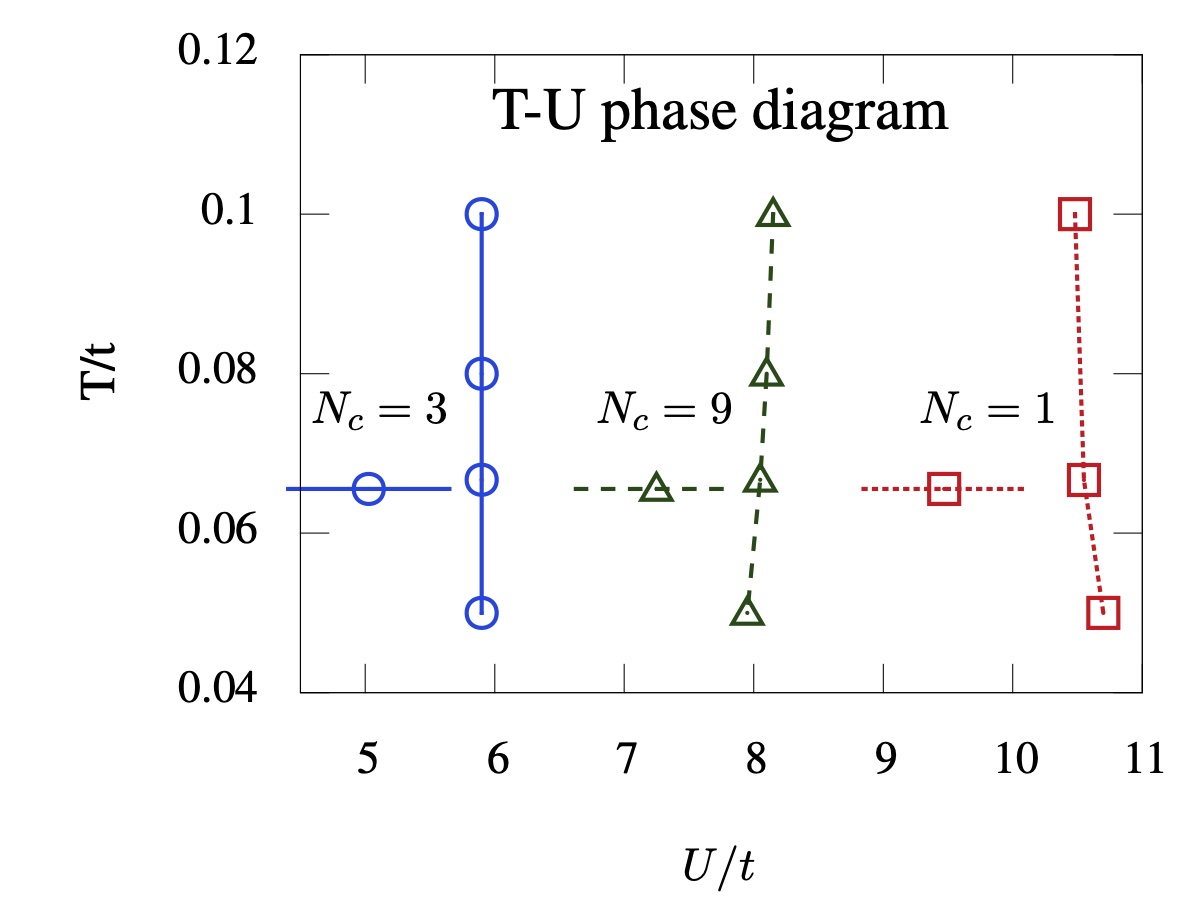}
\caption{(Top) Single-particle Green's function $G_{K}(\tau)$ at imaginary-time $\tau = 1/2T$ of the isotropic triangular lattice as functions of $U$. 
The main plot shows $G_{K}(\tau=1/2T)$ at two selected temperatures $T  = 1/15$ and $T  = 1/20$. $K$ is chosen as the momentum patch where Fermi surface resides.
(Bottom) The T-U phase diagram summarized on all four temperatures studied in this work and at three different size of clusters.}
\label{Fig:MIT}
\end{figure}

With the $N_{c}=9$ site DCA calculation, we want to first answer this question. 
To study the MIT, we examine the interacting Green's function at the momentum patches colored in blue in Fig.~\ref{Fig:cluster_BZ}(b), where the FS of the metallic phase resides. 
In Fig.~\ref{Fig:MIT}, we show the imaginary-time single-particle Green's function at $\tau=1/2T$, where $T$ is temperature, at these momentum patches for three clusters with different size, i.e. $N_{c}=1, 3$, and 9.
Let's focus on the $N_{c}=9$ calculations.
When approaching the insulating phase from the metallic side, $G_{k}(1/2T)$ decreases slowly from a finite value and logarithmically drops to zero when crossing the transition boundary. 
Thus, it works as a probe of the MIT. 
In the top plot of Fig.~\ref{Fig:MIT}, we show the evolution of $G_{K}(\tau)$ for cluster momentum $K$ at $\#2-\#7$ patches and $\tau=t/2T$ at two different temperatures. 
For both temperatures, we observe a clear suppression of $G_{K}(1/2T)$ at $U_{c} \sim 8$ eV.
Uc obtained in our symmetry-invariant 9-site DCA calculations agrees well with the published results~\cite{doi:10.1143/JPSJ.76.074719, PhysRevLett.100.136402, PhysRevLett.100.076402, PhysRevLett.101.166403, PhysRevB.77.214505, PhysRevLett.110.206402, PhysRevB.89.195108, PhysRevB.89.161118, PhysRevB.91.155101, PhysRevB.95.075124, PhysRevB.96.205130, PhysRevResearch.2.013295}.
Furthermore, with the decrease of temperature, $U_{c}$ becomes smaller. 
This behavior is similar to the MIT on square lattice~\cite{PhysRevLett.101.186403}, as DCA makes no approximation on the local charge fluctuations within the cluster as in the cellular DMFT.
Summarizing the results on all four temperatures we studied in this work, we show the transition boundary in the bottom plot of Fig.~\ref{Fig:MIT}. 
The back-turning of MIT boundary is obvious, indicating the importance of the nonlocal correlation effect.
Compared to the square lattice study, both $U_{c}$ and the degree of back-turning are smaller due to the geometrical frustration. 
However, the lack of nonlocal charge fluctuations leads an increasing $U_{c}$ with the decrease of temperature in single-impurity DMFT ($N_{c}=1$), see the right most plot in each figure. 
The $N_{c}=3$ DCA cluster partially corrects the MIT boundary, leading to a constant value of $U_{c}$ at all temperatures studied. 
Our previous DF calculation is consistent with this conclusion, but with a larger $U_{c}$ value due to the different approximations~\cite{PhysRevB.89.161118}. 
Note that, in the $N_{c}=3$ DCA cluster, each pair of sites is connected by hopping inside the cluster as well as hopping through the periodic boundary. Thus, the $N_{c} = 3$ cluster is subjected to a stronger boundary effect, as the $N_c= 2, 4$ DCA clusters for the square lattice.
The $N_{c}=9$ DCA calculations incorporate more nonlocal correlations and are less affected by the boundary effect. 
We, thus, believe that the predicted back-turning of the MIT boundary is an intrinsic character of the triangular lattice.

\begin{figure}[htbp]
\centering
\includegraphics[width=\linewidth]{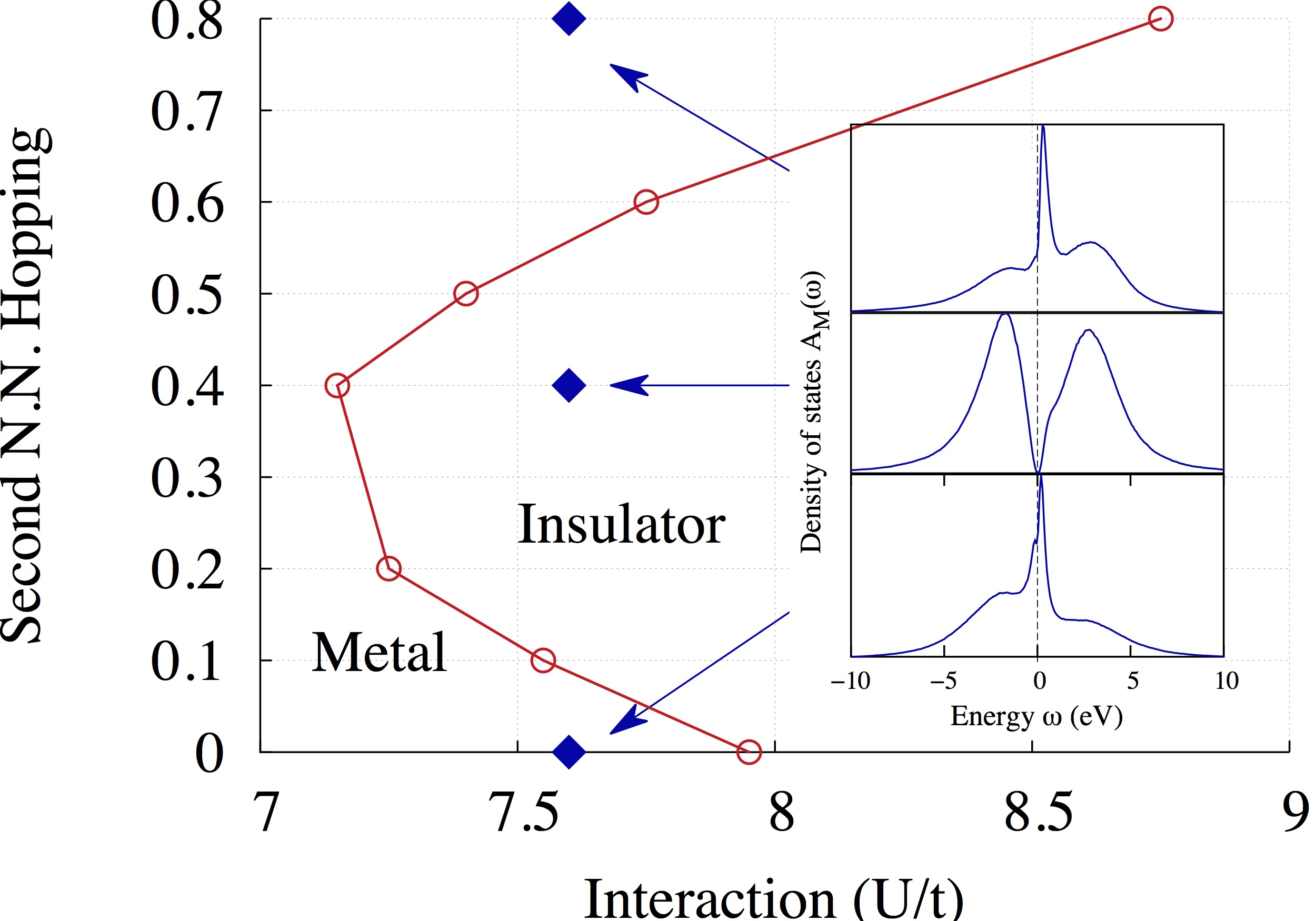}
\caption{The critical value $U_{2}$ for the Metal-Insulator transition (red circled line) does not monotonically depend on the second-N.N. hopping amplitude. For a fixed value of interaction, Metal-Insulator-Metal transition can then be observed. The insets show the local density of state at $U=7.6$ eV and $t^{\prime}=0.0, 0.4, 0.8$. }
\label{Fig:U-tp}
\end{figure}

Now we further include the second N.N. hopping $t^{\prime}$ and examine the MIT boundary at fixed temperature $T = 0.05$. 
We show the estimated $U_{c}$ for different values of $t^{\prime}$. 
It is very interesting to observe that the MIT boundary does not monotonically vary with the change of $t^{\prime}$. 
Increasing $t^{\prime}$, $U_{c}$ rather shows a decreasing followed by an increasing behavior as displayed in Fig.~\ref{Fig:U-tp}. 
Varying $t^{\prime}$ from 0 to 1 at fixed $7.2 ~\mbox{eV} <U < 7.95~\mbox{eV}$, one will first observe a metallic state with FS at small $t^{\prime}$, then the FS disappears for some intermediate values of $t^{\prime}$, and further increasing $t^{\prime}$ leads to the appearance of the FS again.  
Thus, the second N.N. hopping in triangular lattice results in a metal-insulator-metal transition - a reentrance of the MIT. 
This behavior has not been observed in nonfrustrated lattice, where only one type of magnetic correlations dominates. 
As shown below, we will see that the reentrance of MIT strongly connects to the competition of two different magnetisms.  

The reentrance of MIT is confirmed by the local density of states calculated at $U=7.6$ eV shown inside Fig.~\ref{Fig:U-tp}. 
The local density of states correspond to the imaginary-part of the momentum-averaged lattice Green's function. 
We calculated the lattice Green's function in Matsubara frequency space, and transformed it to real frequency by using the stochastic analytical continuation~\cite{beach2004identifying}. 
When $t^{\prime}/t=0$, the isotropic triangular lattice is a metal with a quasiparticle peak at the Fermi level. 
At $t^{\prime}/t=0.4$, the FS completely vanishes. The local density of states show no weight at the Fermi level, which appears again when $t^{\prime}/t$ further increases to 0.8. 

The reentrant MIT in our paramagnetic DCA calculations is closely related to the competition of two different magnetic correlations. 
Although the magnetic fluctuations are suppressed in the single-particle level as we averaged the Green's function in each DCA iteration, the two-particle magnetic fluctuations are still present.  
At smaller or larger $t'/t$, each of the two magnetic correlations dominates. While, at intermediate $t'/t$ they strongly compete. 

\section{Magnetic competition}
\label{Sec:magnetic}
To better understand the magnetic competition, we examine the instability of the magnetic channel by employing the Bethe-Salpeter equation (BSE), whose eigen-equation reads:
\begin{equation}\label{BSE_eigen}
-\frac{T}{N}\sum_{K^{\prime}, \tilde{k}}\Gamma_{Q}^{K, K^{\prime}}G(K^{\prime}+\tilde{k})G(K^{\prime}+Q+\tilde{k})\Psi_{Q}(K^{\prime}) = \lambda_{Q}\Psi_{Q}(K)\;.
\end{equation}
Here $K$, $K^{\prime}$ and $Q$ are the joint variables containing the cluster momenta and the Matsubara frequencies. As we work in the paramagnetic phase, we drop off the spin dependence of the cluster Green's function for simplicity. 
$\Gamma_{Q}(K, K^{\prime})$ is the cluster vertex function calculated from the two-particle cluster Green's function $\chi_{Q}(K, K^{\prime})$ as
\begin{equation}
\Gamma_{Q}^{K, K^{\prime}} = \chi_{0, Q}^{-1}(K, K^{\prime}) - \chi_{Q}^{-1}(K, K^{\prime})\;,
\end{equation}
with $\chi_{0, Q}(K, K^{\prime})$ being the cluster bubble susceptibility. $\chi_{Q}(K, K^{\prime})$ is measured in the last iteration of the DMFT self-consistent loop after convergence is achieved. 
We plot the leading eigenvalue $\lambda_{Q}$ at $Q=\mathbf{K}$ and $Q=\mathbf{M}$, see Fig.~\ref{Fig:cluster_BZ} for more details. 
When the leading eigenvalue approaches 1, the magnetic channel at the corresponding $Q$ point will become divergent signaling the breakdown of the convergence in BSE, which indicates the instability of the paramagnetic solution. 
As a result, a spontaneous phase transition would occur towards a magnetic phase with the magnetic wave-vector $Q$. 
Thus, by comparing the leading eigenvalue $\lambda_{Q}$ we can know if a magnetic instability is going to develop, and, correspondingly, the type of magnetic correlations.  

\begin{figure}[htbp]
\centering
\includegraphics[width=0.9\linewidth]{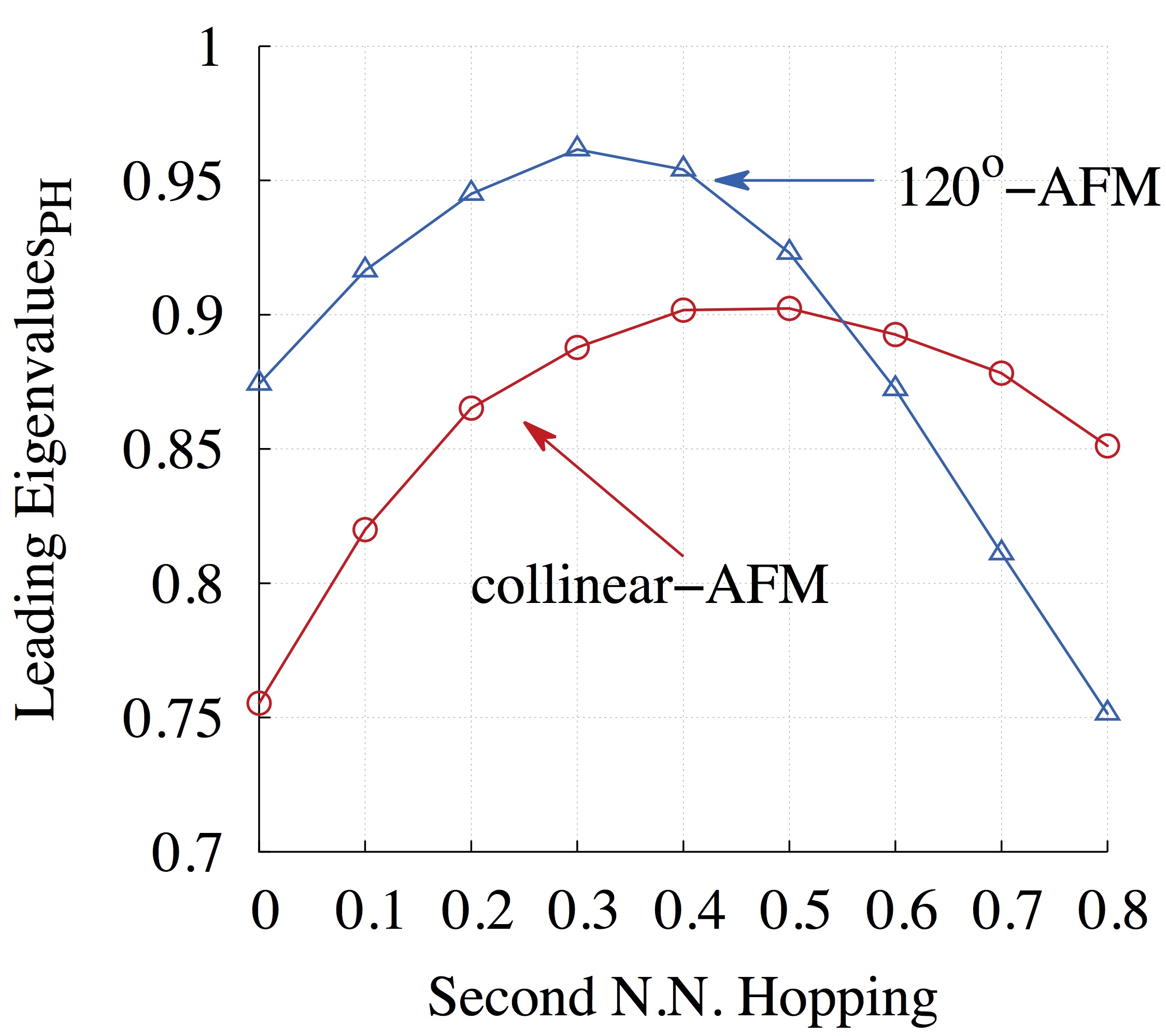}
\caption{The leading eigenvalue of the magnetic channel with $U=7$ eV at two-different high-symmetry momenta, {\it i.e.} $\mathbf{K}$ and $\mathbf{M}$, whose vectors correspond to the magnetic wave vector of the $120^{\circ}$-AFM and the collinear-AFM. When $t^{\prime} >0.55$, the collinear-AFM becomes the leading instability of the system. }
\label{Fig:Eigen}
\end{figure}
Figure~\ref{Fig:Eigen} shows the leading eigenvalue in the magnetic channel with $U=7$ at $\#2-\#7$  and $\#8-\#9$ momentum patches, respectively. 
They correspond to the collinear AFM and  $120^{\circ}$-AFM correlations.
At the $\#1$ momentum patch, the eigenvalue is smaller than the other patches and is not shown.
As expected, without second N.N. hopping, the leading magnetic eigenvalue at patches $\#8-\#9$ wins consistent with the $120^{\circ}$ AFM in isotropic triangular lattice.
However, it quickly drops down when $t^{\prime}$ becomes larger than 0.3. 

The leading eigenvalues shown in Fig.~\ref{Fig:Eigen} further approaches one when we reduce temperature. We note that this does not necessarily correspond to the establishment of a long-rang magnetic order, which is strictly prohibited at 2D in systems with continuously rotational symmetry~\cite{PhysRevLett.17.1133, PhysRevLett.17.1307}. There are two reasons for the finite-temperature magnetic ordering observed in our calculations. First, DCA is a cluster extension of DMFT, which partially incorporates the nonlocal correlation effect. Any longer-range correlation beyond the cluster scope is still treated as mean-field. As in most of the mean-field calculations, DCA would still predict a finite transition temperature towards magnetic ordering, which will be gradually suppressed with the increase of cluster size. Second, the break-down of the BSE in a finite-size cluster only indicates that the magnetic correlation length exceeds the cluster size. It may not correspond to a true long-range order. Here, one should understand Fig.~\ref{Fig:Eigen} as the competition of two different magnetic correlations. Whether they will lead to a true magnetic ordering cannot be unbiasedly answered by our finite-size cluster study.

\begin{figure}[htbp]
\centering
\includegraphics[width=0.9\linewidth]{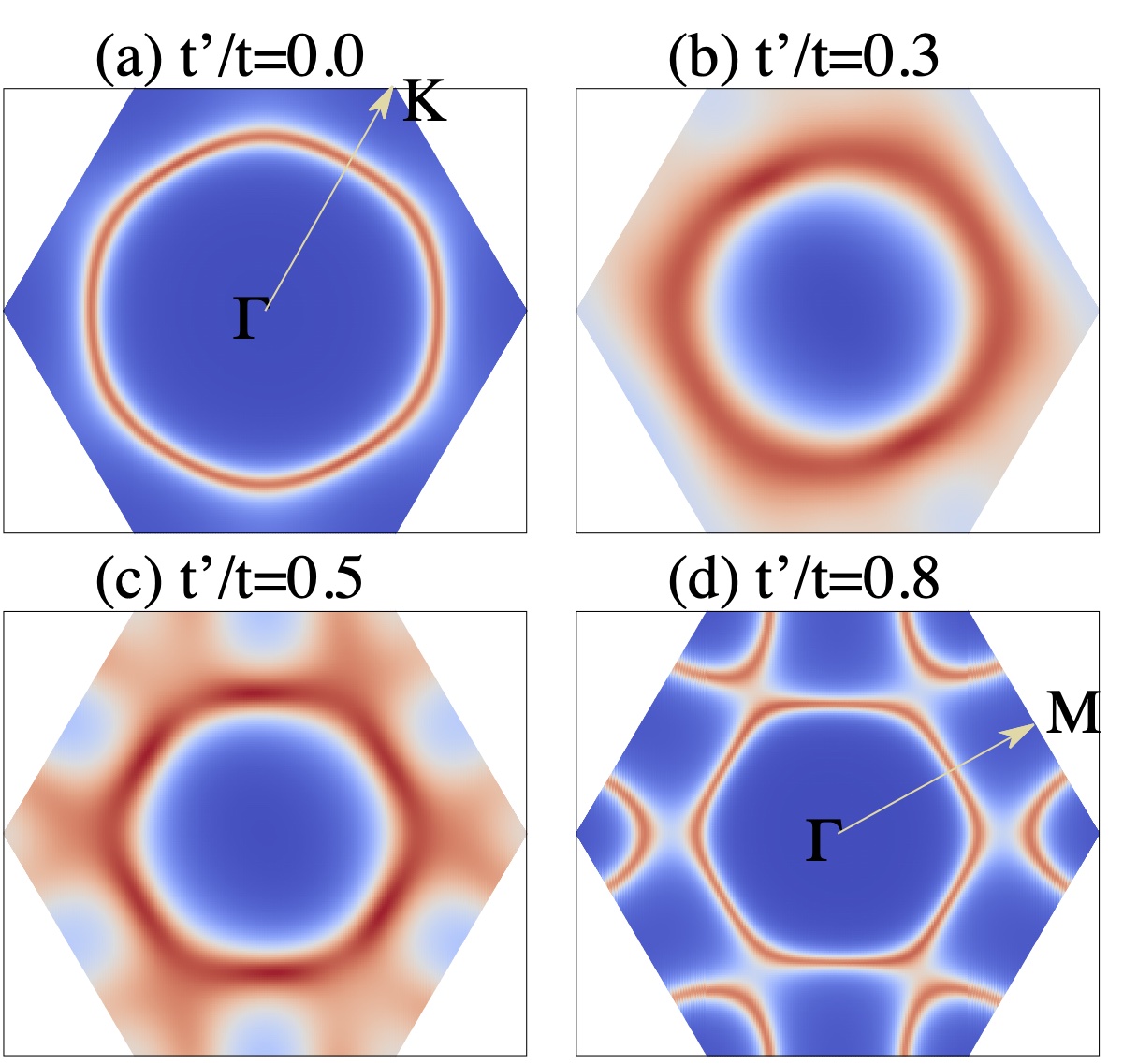}
\caption{The Lifshitz transition of the FS topology as a function of $t^{\prime} $ at $U =7$. From blue to red, the intensity increases.}
\label{Fig:FS}
\end{figure}
The magnetic competition can also be understood from the topology of the quasiparticle FS. 
We found that, in isotropic triangular lattice with second N.N. hopping, the FS of the correlated electrons shows distinct topology at smaller and larger $t^{\prime}$. 
And, in both cases, the FS displays nesting shapes with different nesting vectors. 
To examine the FS, we need to calculate the momentum-dependent single-particle Green's function $G_{k}(\omega)$. 
\begin{equation}
A(k, \omega=0) = -\frac{1}{\pi}\Im\frac{1}{i\delta - \epsilon_{k} - \Sigma_{k}(\omega=0)}\;.
\end{equation}
In DCA calculations with a finite number of cluster size, instead of $\Sigma_{k}(\omega)$, which is a smooth function in the entire BZ, one has step wise $\Sigma_{K}(\omega)$ that is discontinuous when crossing the momentum patch boundary. 
Thus, if calculated with $\Sigma_{K}(\omega)$, $A(k, \omega)$ would be discontinuous as well. 
To get a smooth $\Sigma_{k}(\omega)$ in momentum space, we adopt the $\Sigma$-periodization scheme~\cite{PhysRevLett.87.186401}. 
\begin{equation}
\Sigma_k(\omega) = \frac{1}{N_{c}} \sum_{i,j}\Sigma_{i,j}(\omega)e^{-i\mathbf{k}\cdot(\mathbf{r}_{i}-\mathbf{r}_{j})}\;,
\end{equation}
where $\Sigma_{i,j}(\omega)$ is the cluster self-energy with $i, j$ running over the limited number of cluster sites. 
Figure~\ref{Fig:FS} shows the spectral function over the first BZ as an intensity plot with $U = 7$ eV. The intensity increases in color from blue to red. 
Four representative $t^{\prime}$ are taken in these calculations. 
The FS of the isotropic triangular lattice in Fig.~\ref{Fig:FS}(a) at half-filling shows a hexagonal shape with the different pieces of the FS connected by a fixed wave vector that is equal to the vector from $\mathbf{\Gamma}$ to $\mathbf{K}$. 
Thus, the spin susceptibility at $\mathbf{q}=\mathbf{K} - \mathbf{\Gamma}$ will be enhanced, yielding the tendency toward a $120^{\circ}$ AFM spontaneous symmetry breaking. 
In contrast, at $t^{\prime}=0.8$ [Fig.~\ref{Fig:FS}(d)], the FS shrinks to a smaller hexagon.
The nesting wave vector becomes equivalent to vector $\mathbf{M}$, which is the magnetic wave vector of row-wise collinear AFM.  
In the intermediate values of $t^{\prime}$, the FS smoothly interpolates between those in Fig.~\ref{Fig:FS}(a) and (d). ƒ
The FS evolution is highly consistent with that of the magnetic correlations shown in Fig.~\ref{Fig:Eigen}.  
As either smaller or larger $t^{\prime}$, the dominant magnetic correlations can be interpreted from the corresponding single-particle FS topology, 
while, in the intermediate $t^{\prime}$, the FS nesting vectors do not correspond to a commensurate lattice vector. 
The corresponding magnetic correlation is a superposition of the  $120^{\circ}$ AFM and the row-wise collinear AFM, reflecting the strong competition between these two magnetic correlations.
We note that, at $t^{\prime}=0$ and 0.8, and $U=7$, the system is deeper in the metallic phase as compared to the case of $t^{\prime}=0.3$ and $0.5$.
Their quasiparitcle FS are well defined in Fig.~\ref{Fig:FS}(a) and (d), which resemble the non-interacting ones shown in Fig.~\ref{Fig:cluster_BZ}(d).
However, close to the MIT boundary, the self-energy blurs the single-particle spectra at $t^{\prime}=0.3$ and $0.5$, indicating a stronger correlation effect in Figs.~\ref{Fig:FS}(b) and (c) although the same $U $ was taken as in the cases of Figs.~\ref{Fig:FS}(a) and (d).

\section{Discussion and Conclusions} 
\label{Sec:conclusion}
In this work, we systematically studied the isotropic triangular lattice with the second N.N. hopping, which is a realistic model for various correlated triangular surface systems.  
We found that, despite the geometrical frustration, the nonlocality in triangular lattice still plays an important role. 
The MIT boundary shows a back-turning shape with the decrease of temperature (see Fig.~\ref{Fig:MIT}), similar as in the nonfrustrated square lattice, i.e., a behavior that has not been studied before to our knowledge.
Our calculations strongly indicate the insufficiency of the single-site DMFT calculation in studying the phase boundaries of such model.
We found that the competition of $t^{\prime}$ and $t$ results in a reentrant shape of the MIT (see Fig.~\ref{Fig:U-tp}).
At intermediate values of $t^{\prime}$, the insulating phase can be stabilized at smaller $U$ as contrast to the $t$- or $t^{\prime}$-dominant parameter regime.  
The dominant $120^{\circ}$ AFM and the row-wise collinear-AFM at smaller and larger $t^{\prime}$ respectively become strongly competitive in the intermediate regime of $t^{\prime}$. 
Thus, a nonmagnetic insulating phase is highly anticipated in this regime at zero-temperature~\cite{PhysRevB.95.075124}. 
The second N.N. hopping, thus, provides a more reliable tuning parameter in triangular lattice to achieve nonmagnetic insulating phase, which is highly feasible and relevant in real material systems. 
The evolution of the magnetic correlations can be consistently explained by the Fermi surface topology of the quasiparticle. 
The competition of $t^{\prime}$ and $t$ results in a more featureless FS in the intermediate $t^{\prime}$ regime, while, for the small and large $t^{\prime}$ the FS shows nesting shape with well-defined nesting vectors consistent with the magnetic correlations. 
Our finite-temperature study provides firm evidence that the magnetic correlations are strongly sensitive to the presence of longer-range hopping~\cite{nat.comm_2013,PhysRevB.90.125439}.    

\section{Acknowledgements} 
This work was supported by the National Natural Science Foundation of China under Grant No. 11874263.
X.W. was supported by the National Program on Key Research Project (Grant No. 2016YFA0300501) and by the National Natural Science Foundation of China (Grants No. 11574200 and No. 11974244).
The authors gratefully acknowledge the Gauss Centre for Supercomputing for funding this project by providing computing time on the GCS Supercomputer SuperMUC at Leibniz Supercomputing Centre (LRZ).
Part of the calculations were performed at the HPC Platform of ShanghaiTech University Library and Information Services, and at School of Physical Science and Technology.

\bibliographystyle{apsrev4-1}
\bibliography{ref}

\end{document}